\documentclass[aps, 10pt, amsmath, twocolumn, superscriptaddress, floatfix]{revtex4-1}
\usepackage[plainpages=false,colorlinks=true,linkcolor=blue, citecolor=blue, urlcolor=blue]{hyperref}
\usepackage[english]{babel}
\usepackage{graphicx}
\usepackage{array}
\usepackage[charter]{mathdesign}
\usepackage{bm}
\usepackage[svgnames]{xcolor}

\usepackage{array}

\newcolumntype{C}{>{$\displaystyle}c<{$}}
\newcolumntype{L}{>{$\displaystyle}l<{$}}
\newcolumntype{R}{>{$\displaystyle}r<{$}}

\renewcommand\a{\alpha}
\renewcommand\b{\beta}

\newcommand\xv{\mathbf{x}}
\newcommand\yv{\mathbf{y}}

\newcommand\rv{\mathbf{r}}
\newcommand\tv{\mathbf{t}}
\newcommand\kv{\mathbf{k}}

\newcommand\Gv{\mathbf{G}} 

\newcommand\Vv{\mathbf{V}}
\newcommand\kvt{\mathbf{\tilde k}}

\newcommand\Sigmav{\bm{\Sigma}}
\newcommand\epsilonv{\bm{\epsilon}}
\newcommand\thetav{\bm{\theta}}
\newcommand\Gammav{\bm{\Gamma}}
\newcommand{\dg}{\dagger}
\newcommand{\fdg}{{\phantom{\dagger}}}
\newcommand{\up}{{\uparrow}}
\newcommand{\down}{\downarrow}
\newcommand\ket[1]{|#1\rangle}
\newcommand\bra[1]{\langle#1|}

\newcommand\unit{\mathbf{1}}

\begin{document}
\title{Doping phase diagram of a Hubbard model for twisted bilayer cuprates}
\author{Xiancong Lu}
\affiliation{Department of Physics, Xiamen University, Xiamen 361005, China}
\author{D. S\'en\'echal}
\affiliation{D\'epartement de physique and Institut quantique, Universit\'e de Sherbrooke, Sherbrooke, Qu\'ebec, Canada J1K 2R1}

\begin{abstract}
We study the twisted Hubbard model of a cuprate bilayer at a fixed twist angle $\theta=53.13^{\circ}$ using the variational cluster approximation, a method that treats short-range dynamical correlations exactly.
At intermediate interlayer tunneling, the phase difference $\phi$ between the $d$-wave order parameters of two layers is $\pi$ in the overdoped regime, while it is zero in the underdoped regime, close to the Mott phase.
At strong interlayer tunneling, we observe a clear time-reversal symmetry breaking phase near optimal doping, in which the phase difference $\phi$ changes continuously from 0 to $\pi$. However, this phase has trivial topology. We also apply a cluster extension of dynamical mean field theory to the same problem, but fail to detect a time-reversal breaking phase with that method.
\end{abstract}
\maketitle
\section{Introduction}

The experimental discovery of correlated insulators and unconventional superconductivity in twisted bilayer graphene (TBG) \cite{ca.fa.18a,ca.fa.18b} has opened up the new field of twistronics \cite{an.ma.21,ca.ma.17}. 
By twisting two graphene sheets by a small relative angle, a long-period moir\'e pattern forms in the bilayer. 
At special \textit{magic} angles, the moir\'e band structure of TBG exhibits isolated flat bands near charge neutrality \cite{bi.ma.11,sa.pe.07,ta.kr.19}, which lead to a variety of strongly correlated phenomena. 
Following this discovery, various twisted van der Waals heterostructures have been constructed and investigated \cite{ke.cl.21}, including transition metal dichalcogenides \cite{re.wa.20,ta.li.20,wa.sh.20,hu.wa.21}, double bilayer graphene \cite{li.ha.20,ca.le.20,zh.zh.21}, and trilayer graphene \cite{ch.ji.19,ch.sh.19,ch.sh.20,pa.ca.21}.

Recently, twistronics concepts have been extended to high-temperature superconductors~\cite{ca.tu.21,vo.wi.20}, which are strongly correlated materials by themselves.
This was motivated by the experimental realization of two-dimensional (2D) monolayer
Bi$_2$Sr$_2$CaCu$_2$O$_{8+\delta}$ (Bi2212), whose transition temperature is shown to be very close to that of bulk samples~\cite{yu.ma.19,zh.po.19}.  
It is theoretically predicted that, at large twist angles (close to $45^\circ$), a fully gapped $d+id$ superconducting phase emerges, which spontaneously breaks time-reversal symmetry (TRS) and is topologically nontrivial~\cite{ca.tu.21}.
This TRS breaking superconducting phase is also predicted to be stable at small twist angle, due to the strong renormalization of Bogoliubov-de Gennes (BdG) quasiparticles near the nodes \cite{vo.wi.20}.
In order to determine the pairing symmetry of cuprate superconductors, $c$-axis twisted Josephson junctions, formed by stacking two Bi2212 crystals along the $c$-axis, have been realized~\cite{li.ts.99,ta.ha.02,la.or.04,klem.05,ya.qi.18,zh.li.21}.
However, most experimental works did not observe the angular dependence of the Josephson current \cite{li.ts.99,ta.ha.02,zh.li.21}.
Owed to the novel technique of van der Waals stacking, high-quality twisted Bi2212 Josephson junctions with an atomically sharp interface have been successfully fabricated recently~\cite{zh.li.21,zh.po.21}.

Previous theoretical work on twisted bilayer cuprates are mainly based on Bogoliubov-de-Gennes mean-field theory \cite{ya.qi.18,ca.tu.21,vo.wi.20,tu.pl.21,vo.zh.21}, which does not take into account the effects of strong correlations.
To overcome this, a twisted $t$-$J$ model of cuprates has been proposed and studied within slave-boson mean-field theory \cite{so.zh.21}, in which a topological-trivial time-reversal symmetry breaking superconductor is also found, but within a small range of twist angles around $45^{\circ}$, questioning the possibility of topological superconductors in this region.
In spite of this work, the stability of the novel superconducting phases against doping has not been fully addressed before in the literature.
In this paper, we will numerically study the twisted Hubbard model of bilayer cuprates using the variational cluster approach (VCA) and cluster dynamical mean field theory (CDMFT).
These approaches have been successfully used in the past to study high-temperature superconductors and the Hubbard model at intermediate coupling is arguably a better representation of these materials.
We will focus on a fixed twisted angle $\theta=53.13^{\circ}$, at which these cluster methods are easily applicable, and investigate the superconducting phase diagram as a function of doping for two different sets of interlayer tunneling.

This paper is organized as follows.
In Sec.~II, we introduce the Hubbard model for the twisted bilayer.
In Sec.~III, we review the variational cluster approximation (VCA) and present our main results obtained from this method, \textit{e.g.}, the phase diagram of bilayer as a function of hole doping.
In Sect.~IV, we present the corresponding results from cluster dynamical mean field theory (CDMFT).
	
\section{Model}\label{sec:intro}
\subsection{Hamiltonian}

We assume that each of the two layers of the system can be described by the one-band Hubbard model (the sites correspond to the location of copper atoms).
The bilayer is then described by the following tight-binding Hubbard model:\cite{ca.tu.21,so.zh.21}
\begin{equation}\label{eq:Htot}
H = H^{(1)} + H^{(2)}  + H_\perp. 
\end{equation}
where the intra-layer Hamiltonian $H^{(\ell)}$ is
\begin{equation}\label{eq:Hintra}
H^{(\ell)} = \sum_{\rv,\rv'\in\ell, \sigma} t_{\rv\rv'} c_{\rv,\ell,\sigma}^\dg c_{\rv',\ell,\sigma}
+ U \sum_{\rv} n_{\rv,\ell,\uparrow} n_{\rv,\ell,\downarrow} - \mu\sum_{\rv,\sigma} n_{\rv,\ell,\sigma},
\end{equation}
where $c_{\rv,\ell,\sigma}(c_{\rv,\ell,\sigma}^\dagger)$ is the annihilation (creation) operator of an electron at site $\rv$ on layer $\ell$ ($\ell=1,2$) with spin $\sigma=\uparrow,\downarrow$, and $n_{\rv,\ell,\sigma}$ is the associated number density operator.
The labels $\rv,\rv'$ run over the possible sites of a square lattice (each layer has it own).
We will keep nearest-neighbor ($t$) and next-nearest-neighbor ($t'$) hopping terms only, so that the dispersion relation on a square lattice is 
$\varepsilon(\kv) = -2t(\cos k_x + \cos k_y) +4t'\cos k_x\cos k_y - \mu$.
Only on-site interactions are considered here.
For Bi2212, the nearest-neighbor hopping is $t=126$meV \cite{ma.sa.05}.
In the remainder of this paper, we set $t$ as the energy unit, and choose the other parameters to be $t'=-0.3$ and $U=8$.

\begin{table}
\begin{tabular}{m{15mm}m{15mm}m{15mm}m{15mm}}
set & $V_1$ & $V_2$ & $V_3$ \\ \hline
I & 0.1 & 0.05 & 0.03 \\
II & 0.4 & 0.2 & 0.012 \\ \hline
\end{tabular}
\caption{The two sets of inter-layer hopping terms used in this work.\label{table:V}}
\end{table}

\begin{figure}[tp]
\includegraphics[scale=0.75]{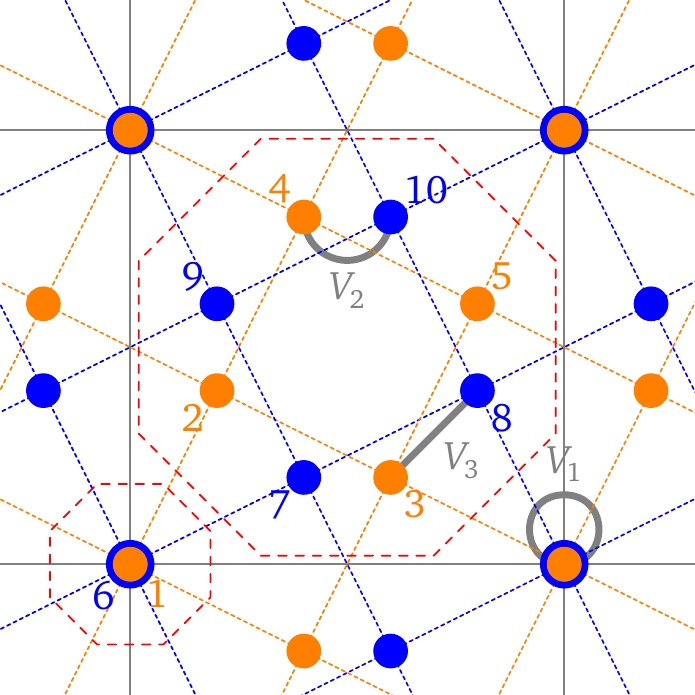}
\caption{(Color online). Unit cell of the bilayer twisted by an angle $\theta=2\arctan\frac12 = 53.13^\circ$.
The ten sites within the unit cell are labeled and their color (orange or blue) indicates the layer.
The three most important interlayer tunneling terms ($V_{1,2,3}$) are illustrated in gray.
(the red dashed enclosures are the clusters used in VCA; see below).
\label{fig:cluster}}
\end{figure}

The interlayer tunneling is represented by
\begin{equation}
H_\perp = \sum_{n=1}^3 V_n\sum_{\langle \rv,\rv'\rangle_{\bot,n},\sigma}  \Big[ c_{\rv,1,\sigma}^\dg c_{\rv',2,\sigma} + \mbox{H.c.} \Big] 
\end{equation}
where the notation $\langle \rv,\rv'\rangle_{\bot,n}$ ($n=1,2,3$) stands for the set of square lattice sites $\rv$ on layer 1 and $\rv'$ on layer 2 such that their projection on the plane are $n^{\rm th}$ neighbors.
This is illustrated on Fig.~\ref{fig:cluster} for $V_1$, $V_2$ and $V_3$.
For instance, $V_1$ is the interlayer tunneling between sites located exactly on top of each other, $V_2$ for sites that are first neighbors when projected on a common plane, etc.
Such an interlayer tunneling model is obviously oversimplified, as it ignores the complexity of the CuO$_2$ layers and of the rare-earth layers that will intervene between the twisted CuO$_2$ layers.
In this work we will use two sets of values for $V_n$, shown in Table~\ref{table:V}.
These values have been chosen heuristically, those of set II being four times larger than those of set I and certainly unrealistic, but necessary in order to unravel TRS breaking, as we will see below.
Given the accepted values of hopping along the $c$-axis in bulk cuprates, even the values of set I are large, and will be referred to as \textit{intermediate tunneling}, whereas those of set II will constitute \textit{strong tunneling}.

\begin{figure}[tp]
\includegraphics[width=0.5\columnwidth]{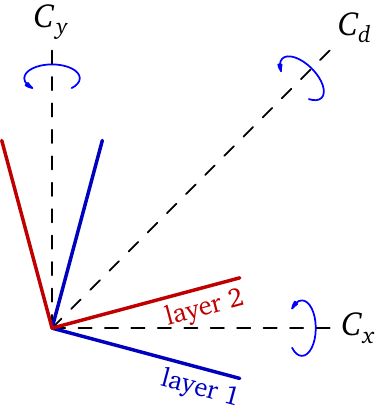}
\caption{(Color online) Symmetries on the bilayer system. The bases of the two layers are indicated in blue and red, respectively.
The rotations $C_x$, $C_y$ and $C_d$ are indicated; the rotations $C_2$ and $C_4$, within each plane, are not.}
\label{fig:D4}
\end{figure}
\begin{table}
\caption{
Character table of $D_4$, with a list of the simplest gap basis functions.
The rotation $C_d'$ is defined about the other diagonal axis, at right angle from $C_d$.
\label{table:D4}
}
\centering\medskip
\begin{tabular}{LRRRRRR}
\hline\hline
& e & 2C_4 & C_2 & C_{x,y} & C_{d,d'} & \mbox{gap functions} \\ 	\hline
A_{1} & 1 & 1 & 1 & 1 & 1      &  1  \\ 
A_{2} & 1 & 1 & 1 & -1 & -1    & \sin k_x\;\sin k_y(\cos k_x-\cos k_y)  \\ 
B_{1} & 1 & -1 & 1 & 1 & -1    & \cos k_x-\cos k_y \\ 
B_{2} & 1 & -1 & 1 & -1 & 1    & \sin k_x\;\sin k_y\\ 
E & 2 & 0 & -2 & 0 & 0         & (\sin k_x,\sin k_y) \\	\hline\hline
\end{tabular}
\end{table}

In order to simplify as much as possible our numerical work, we will restrict our analysis to a twist angle of $\theta=2\arctan\frac12 = 53.13^\circ$.
The unit cell of the twisted bilayer at that angle is illustrated on Fig.~\ref{fig:cluster} and contains ten sites (five per layer).

\subsection{Symmetries}

The bilayer system is invariant under a $\pi/4$ rotation around the $z$ axis (perpendicular to the bilayer plane) and under the $\pi$ rotations $C_x$, $C_y$ and $C_d$ illustrated on Fig.~\ref{fig:D4}, which make up the $D_4$ point group, the same as for an isolated layer.
Possible superconducting gap functions for this system should in principle be classified according to the irreducible representations of $D_4$. 
Table~\ref{table:D4} shows the character table and the simplest gap functions associated with each irreducible representation.
Representations $B_1$ and $B_2$ correspond to what is usually called $d_{x^2-y^2}$ and $d_{xy}$, respectively.
Representations $A_1$ and $A_2$ correspond respectively to $s$-wave (or extended $s$-wave) and $f$-wave, and the two-dimensional representation $E$ would correspond to (triplet) $p$-wave, with basis $(p_x,p_y)$.
Thus, the only possibility of a (pure) chiral representation is $p_x + ip_y$, a triplet state that will not occur in this cuprate system.
We rather expect representations $B_1$ and $B_2$ to be realized here, owing to the $d$-wave character of superconductivity in single layers.
In principle, according to the Landau theory of phase transitions, one of those two should prevail just below $T_c$, but there is always the possibility that, the two states ($B_1$ and $B_2$) being very close in energy, a second phase transition occurs below $T_c$ and a complex combination $d_{x^2-y^2}+id_{xy}$ is present at zero temperature.
This is the scenario anticipated in Ref.~\cite{ca.tu.21} and investigated here.

\begin{figure}[tp]
\includegraphics[scale=0.8]{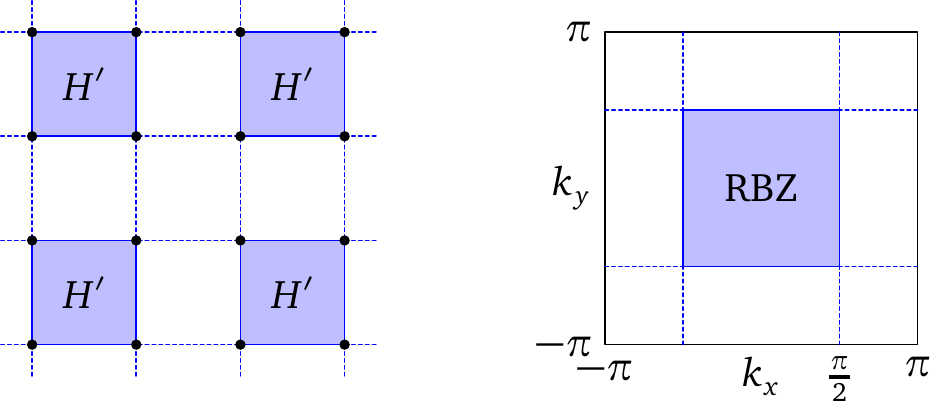}
\caption{(Color online) Schematics of cluster methods. Left panel: the lattice is tiled into identical $2\times2$ clusters with Hamiltonian $H'$. Right panel: the reduced Brillouin zone (RBZ) is then four times smaller than the original Brillouin zone.}
\label{fig:2x2_cluster}
\end{figure}

\section{Results from the Variational Cluster Approximation}

High-temperature superconductors have strong correlations.
There is a limited number of numerical methods that can tackle such systems, and methods based on small clusters of sites embedded into an effective medium are amongst the most successful.
These so-called \textit{quantum cluster methods} are approximation strategies for the electron Green function $G(\kv,\omega)$, by which the electron self-energy $\Sigma$ on the infinite lattice is approximated by that of a small cluster. 
In this work we will apply two of these methods to the bilayer Hamiltonian~\eqref{eq:Htot}.

The first of these methods is the variational cluster approximation (VCA)~\cite{potthoff2003,potthoff2003b,potthoff2014a}.
It is based on a variational principle proposed by Potthoff~\cite{potthoff2003} and can be seen as a variational extension of cluster perturbation theory (CPT)~\cite{senechal2000a, gros1993}.
Let us start by briefly summarizing the latter.
In CPT, the lattice is tiled into identical \textit{clusters}, and the Hamiltonian is written as $H = H' + V$, where $H'$ is the restriction of $H$ to the clusters and $V$ only contains hopping terms between different clusters.
If the model contains $N_b$ bands and each cluster contains $L$ lattice sites, then $LN_b$ must be small enough to allow for an exact numerical solution of $H'$, and the associated one-particle Green function $\Gv_c(\omega)$ on the cluster is a $2LN_b\times 2LN_b$ matrix (the factor of 2 because of spin).
The tiling into clusters defines a superlattice, and the corresponding Brillouin zone is $L$ times smaller than the original Brillouin zone (see Fig.~\ref{fig:2x2_cluster}). 
We call it the reduced Brillouin zone and its wave vectors are noted $\kvt$.
The hopping matrix in $H$ can be expressed as a $2LN_b\times 2LN_b$ matrix $\tv(\kvt)$, a function of $\kvt$, which is the sum of a $\kvt$-independent part $\tv_c$ and of the inter-cluster part $\Vv(\kvt)$ : $\tv(\kvt) = \tv_c + \Vv(\kvt)$.
The self-energy $\Sigmav_c(\omega)$ associated with the cluster Green function $\Gv_c(\omega)$ is thus defined by Dyson's equation on the cluster:
\begin{equation}
\Gv_c^{-1} = \omega - \tv_c - \Sigmav_c(\omega)
\end{equation}
In CPT, the electron self-energy is approximated by that of the restriction $H'$ of the Hamiltonian to the cluster.
In the mixed momentum-cluster site basis, the electron Green function is then given by the following relation:
\begin{equation}\label{eq:Glatt}
	\Gv^{-1}(\kvt,\omega) = \omega - \tv(\kvt) - \Sigmav_c(\omega) = \Gv_c^{-1}(\omega) - \Vv(\kvt)
\end{equation}
We assume here that the chemical potential $\mu$ is included in the hopping matrix $\tv(\kvt)$.

CPT is unable to describe broken symmetry states: it is not a self-consistent approach, nor is it based on a variational principle.
The VCA adds a variational aspect to CPT: the cluster Hamiltonian $H'$ is augmented by a certain number of \textit{Weiss fields}:
\begin{equation}\label{eq:Weiss}
	H' \to  H' + \sum_a h_a {\hat O}_a
\end{equation}
where the operators ${\hat O}_a$ are defined on the cluster only, and possibly represent broken symmetries.
These additional terms are in turn subtracted from $V$, so that the original Hamiltonian $H$ in unaffected.
The values $h_a$ of these Weiss fields are not arbitrary, but set by Potthoff's variational principle:
The following function:
\begin{equation}\label{Potthoff}
\Omega(h_a) =  \Omega' - \int\frac{d\omega}{2\pi}\sum_\kvt \ln\det\left[\unit-\Vv(\kvt)\Gv_c(\omega)\right]
\end{equation}
should be stationary with respect to these fields $h_a$.
In that expression, $\Omega'$ is the ground state energy of the cluster Hamiltonian $H'$ and $\Gv_c(\omega)$ is the electron Green function derived from the cluster Hamiltonian $H'$ that includes the Weiss fields $h_a {\hat O}_a$.

In the problem at hand, it might seem natural to use the 10-site unit cell shown in Fig.~\ref{fig:cluster} as the repeated cluster, especially since 10 sites is an easily manageable size for an exact-diagonalization solver.
However, the set of numbered sites in Fig.~\ref{fig:cluster} does not have the $D_4$ symmetry of the full Hamiltonian, and this complicates the VCA computations. 
We will rather use a slight refinement of the method described above, assuming that the repeated unit is a \textit{supercluster} of 10 sites obtained by assembling an octagonal cluster of 8 sites and a point-like cluster of 2 sites, each delimited by a red dashed line in Fig.~\ref{fig:cluster}. The self-energy of the supercluster is then a direct sum of the self-energies of an 8-site and of a 2-site cluster.
Otherwise, the method is unchanged from the general approach described above.

\begin{figure}[tp]
\includegraphics[width=\columnwidth]{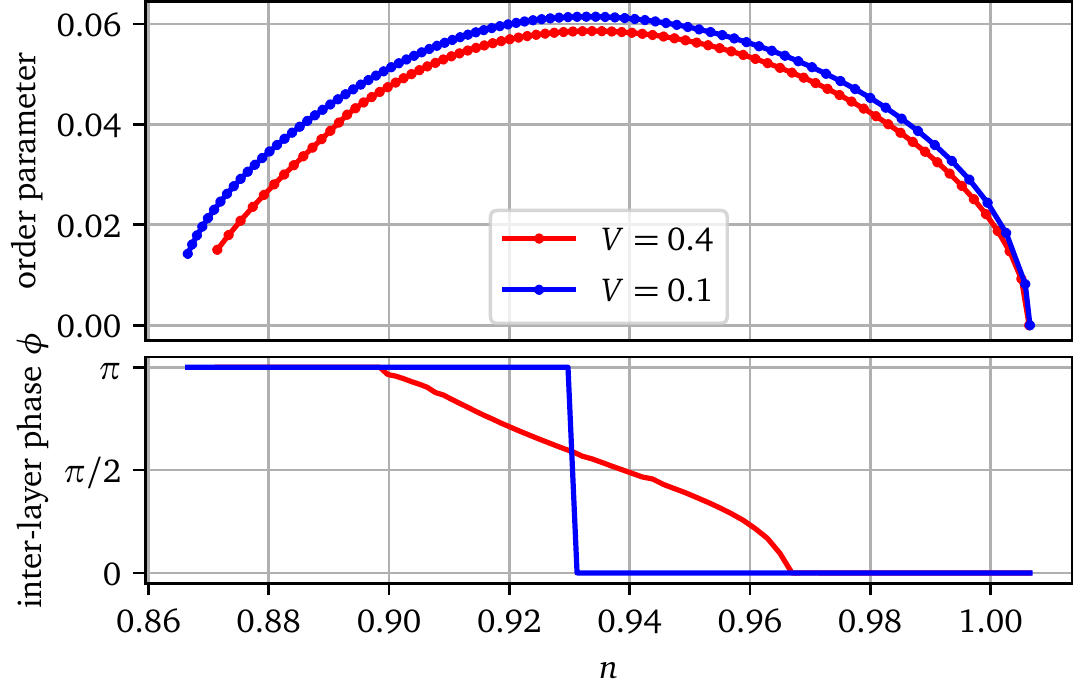}
\caption{(Color online) Top panel: Order parameter as a function of electron density $n$, as obtained in VCA, for both inter-layer hopping sets.
Lower panel: the corresponding relative phase $\phi$ of the order parameters on the two layers. At intermediate inter-layer hopping, the system jumps from $\phi=0$ to $\phi=\pi$ at $n\approx0.93$. At strong inter-layer hopping, the switch is gradual.}
\label{fig:dsc}
\end{figure}

On the octagonal cluster, we will define Weiss fields associated with $d$-wave superconductivity on each layer.
On each layer of the lattice, we can define an operator field that describes $d$-wave superconductivity:
\begin{align}\label{eq:Delta}
\Delta^{(\ell)} =  \sum_{\rv\in\ell} & \Big\{ c_{\rv,\ell,\up}c_{\rv+\xv^{(\ell)},\ell,\down} - c_{\rv,\ell,\down}c_{\rv+\xv^{(\ell)},\ell,\up} 
\notag\\  &- c_{\rv,\ell,\up}c_{\rv+\yv^{(\ell)},\ell,\down} + c_{\rv,\ell,\down}c_{\rv+\yv^{(\ell)},\ell,\up}\Big\} 
\end{align}
where $\xv^{(\ell)}$ and $\yv^{(\ell)}$ are the orthogonal lattice vectors on layer $\ell$.
We can then add the following combinations to the cluster Hamiltonian:
\begin{equation}
H' \to H' + \sum_{\ell=1,2} d^{(\ell)} \Delta^{(\ell)}_c + \mbox{H.c}
\end{equation}
where $\Delta^{(\ell)}_c$ is a restriction to the cluster of the lattice operator \eqref{eq:Delta} and $d^{(\ell)}$ is a complex amplitude.
The real and imaginary parts of $d^{(\ell)}$ are then Weiss fields in the sense of the coefficients $h_a$ of Eq.~\eqref{eq:Weiss}.
Because of overall phase symmetry, one can always assume that $d^{(1)}$ is real, but we must assume in all generality that $d^{(2)}$ is complex.
The complex phase of $d^{(2)}$ is then the relative phase $\phi$ of the superconducting order parameters of the two layers, and a value other than zero or $\pi$ would signal a spontaneous TRS breaking and possible topological properties.

\begin{figure}[tp]
\includegraphics[width=\columnwidth]{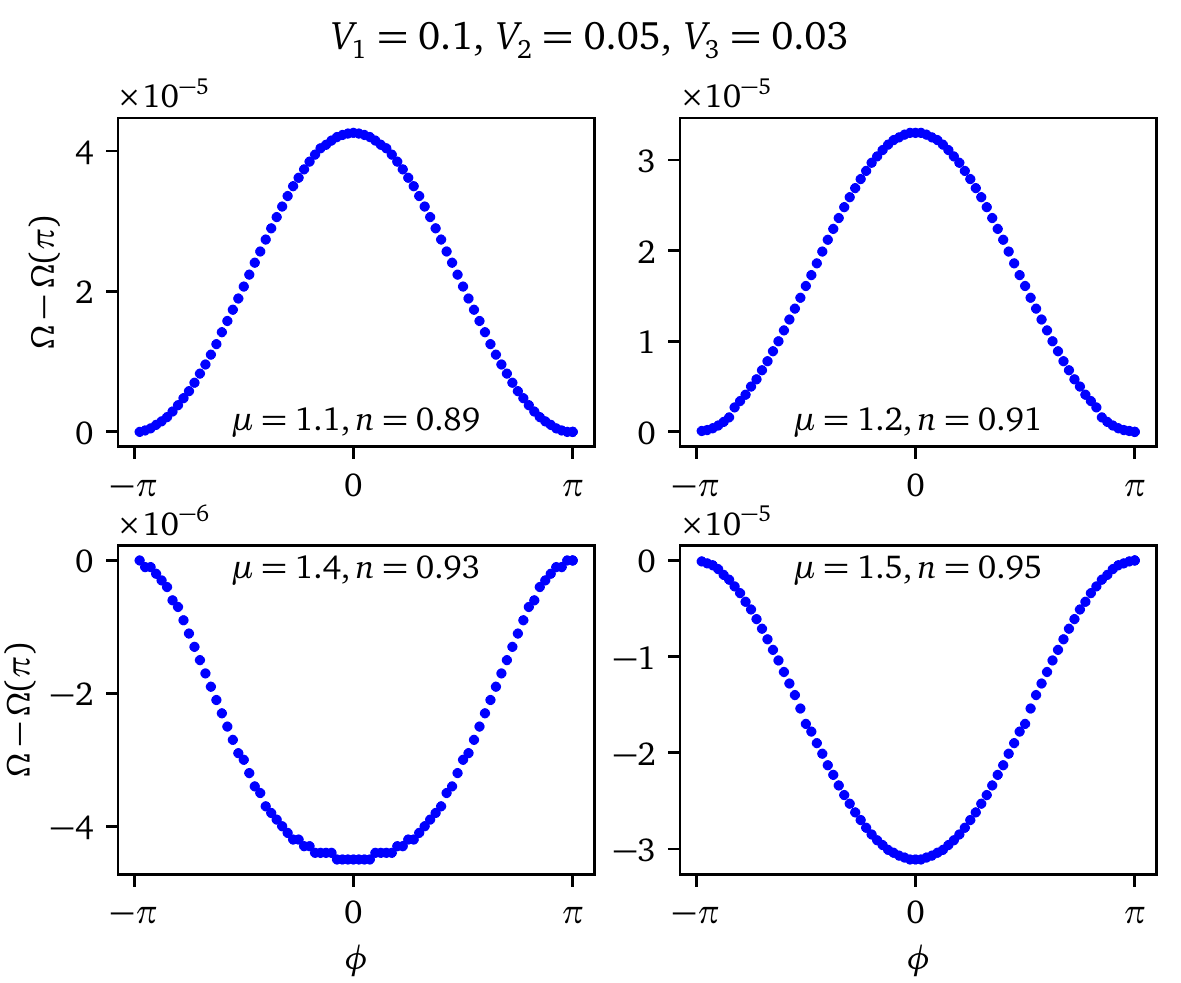}
\includegraphics[width=\columnwidth]{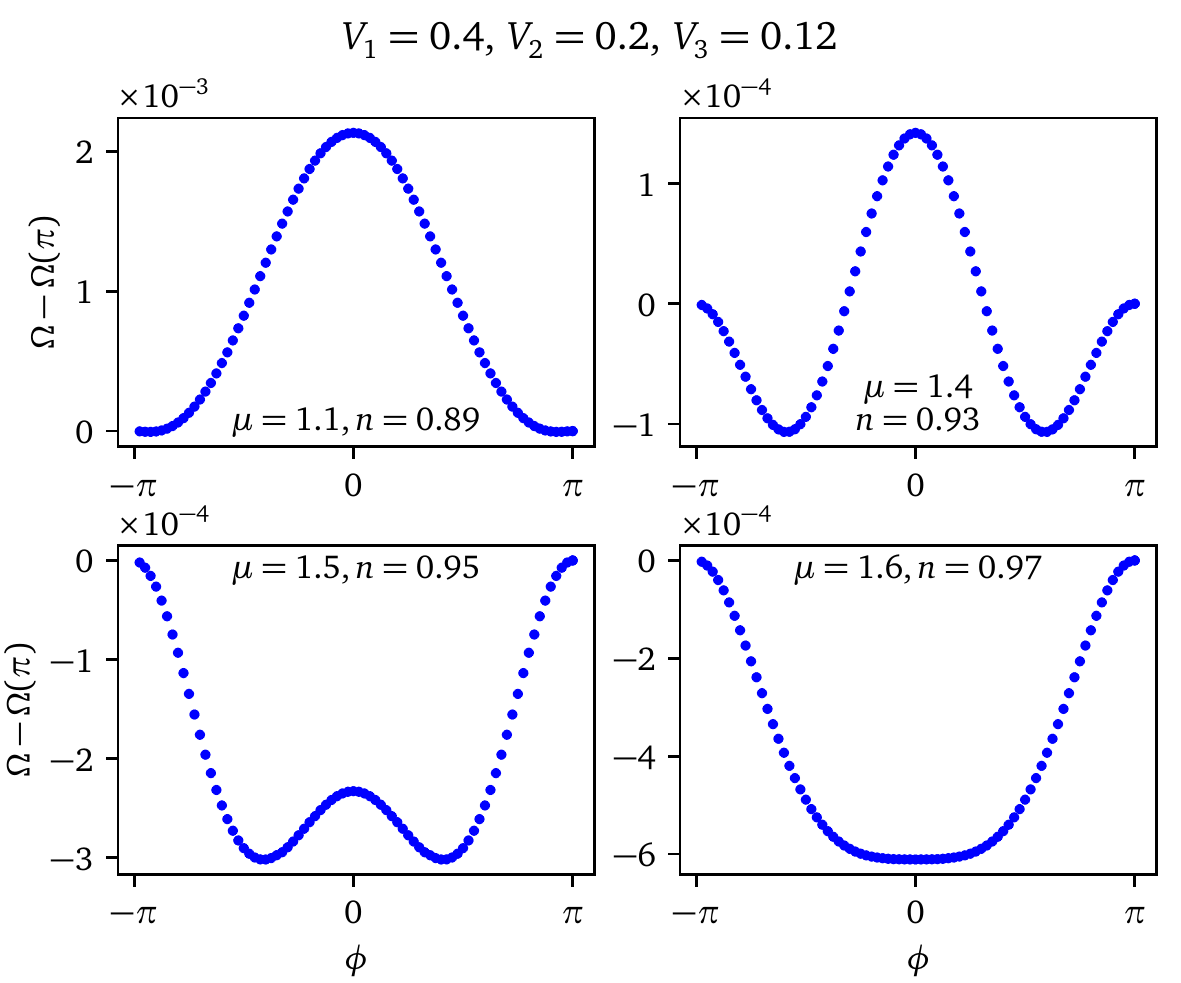}
\caption{(Color online) Potthoff functional as a function of inter-layer phase $\phi$, for different values of the chemical potential, for intermediate (top) and strong (bottom) inter-layer tunneling.}
\label{fig:omega}
\end{figure}

We applied the VCA method on this cluster system, using the two sets of interlayer tunneling defined in Table~\ref{table:V}.
In practice, this means computing the cluster Green function $\Gv_c(\omega)$ repeatedly while adjusting the Weiss fields $d^{(\ell)}$
so as to make the Potthoff functional stationary (in fact, minimum). 
Once the stationary values are found, the Green function \eqref{eq:Glatt} can be used to compute the ground state average of any one-body operator, in particular the order parameter $\Psi^{(\ell)} = \langle \Delta^{(\ell)}\rangle/N$ ($N$ is the number of sites) on each layer.
The electron density $n$ can be likewise computed from the Green function (the chemical potential $\mu$ is the actual control parameter that is varied).

Fig.~\ref{fig:dsc} shows the order parameter $\Psi^{(\ell)}$ as a function of electron density $n$ for hole doping and the two sets of interlayer tunneling (intermediate and strong).
We note the characteristic dome shape that is typically obtained in quantum cluster methods, qualitatively agreeing with the known properties of cuprates.
The electron density computed from the Green function \eqref{eq:Glatt} has some systematic error, as can be seen from the fact that the order parameter vanishes not at $n=1$, as it should from Mott physics, but at $n=1.006$. 
The bottom layer of the figure shows the relative phase of the order parameters
$\Psi^{(2)}$ and $\Psi^{(1)}$ (on the two layers).
At intermediate inter-layer tunneling, this phase is 0 at low doping, which is the signature of the $B_1$ representation of Table~\ref{table:D4}.
Beyond about 7\% doping, this phase switches to $\pi$, a signature of the $B_2$ representation.
There is thus a doping-induced transition of the bilayer superconducting state, which coincides with the passage from underdoped to overdoped, judging by the location of optimal doping on the upper panel of the figure.

For strong interlayer tunneling, the situation is different: an intermediate phase appears in which the relative complex phase of the two order parameters changes continuously from 0 to $\pi$.
This intermediate phase breaks time reversal symmetry and correponds roughly to a
$B_1+iB_2$ state, except that the two components do not have the same amplitude.

A more detailed view of how this is happening from the VCA perspective is shown on Fig.~\ref{fig:omega}. On the top half of the figure, we show the profile of the Potthoff functional~\eqref{Potthoff} as a function of the relative phase $\phi$ of the order parameters on the two layers. 
For $n=0.89$ and $n=0.91$, the minimum is at $\phi=\pm\pi$ ($B_2$ representation).
Near $n=0.93$, the profile changes suddenly to one where the minimum is at $\phi=0$ ($B_1$ representation).
Note that the vertical scale is tiny ($10^{-5}$), in multiples of $t$, which defines the energy unit here.
This means that the energy difference between the two representations $B_1$ and $B_2$ might just be too small to be of consequence experimentally ($\sim 10^{-2}$meV or $\sim 10^{-1}$K in terms of temperature), at an intermediate interlayer tunneling of $V_1=0.1$.

On the bottom half of Fig.~\ref{fig:omega}, the same type of data is shown at strong interlayer tunneling ($V_1=0.4$).
There the transition between $B_2$ and $B_1$ is gradual as the position of the minimum moves continuously from $\phi=\pm\pi$ to $\phi=0$, with a spontaneous breaking of the $\phi\to-\phi$ symmetry.
Even though this TRS breaking state is what we are looking for, such a strong value of interlayer tunneling is unrealistic.

\begin{figure}[tp]
\includegraphics[width=\columnwidth]{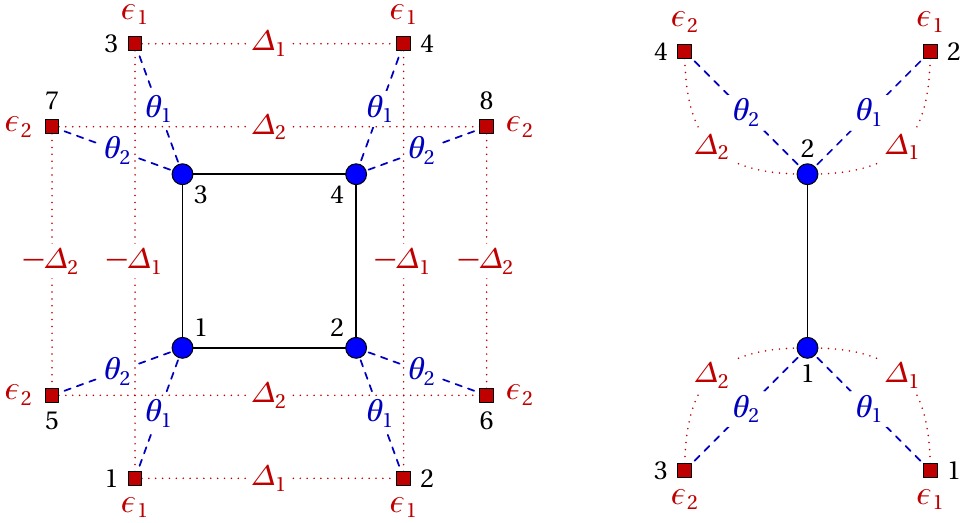}
\caption{(Color online) Impurity models used in CDMFT. On the left: the 4-site cluster used for sites (2,3,4,5) and (7,8,9,10) of the unit cell, as labeled in Fig.~\ref{fig:cluster}. On the right, the 2-site cluster used for sites (1,6).}
\label{fig:bath_simple_SC}
\end{figure}

Does this TRS breaking state have nontrivial topology? In a strongly correlated system, this question may be answered through the properties of the approximate interacting Green function \eqref{eq:Glatt}~\cite{wa.zh.12a, wa.zh.12b}. The key idea is to define a ``topological Hamiltonian'' $h_t(\kv)= - G^{-1}(\kv,\omega=0)$, which can be diagonalized:
\begin{equation}
h_t(\kv) \ket{\alpha,\kv} = \mu_\alpha (\kv) \ket{\alpha,\kv}
\end{equation}
One can then define a generalized Chern number just like in noninteracting systems:
\begin{equation}\label{eq:Chern}
C_1 = \int \frac{d^2k}{2\pi} \; \mathcal{F}_{xy} (\kv)
\quad\quad
\mathcal{F}_{xy} (\kv) = \frac{\partial \mathcal{A}_y}{\partial k_x} - \frac{\partial \mathcal{A}_x}{\partial k_y}
\end{equation}
with the Berry connection
\begin{equation}
\mathcal{A}_j (\kv) = -i \sum_{\mu_\alpha(\kv)<0} \bra{\alpha, \kv} \partial_{k_j} \ket{\alpha,\kv},
\qquad (j = x,y)
\end{equation}
When applying this formula to the TRS states found by VCA, we find the topology to be trivial (the Chern number vanishes). This results from a compensation between different regions of the Brillouin zone, with opposite Berry curvature.

\section{Results from Cluster Dynamical Mean Field Theory}

In order to test the robustness of our predictions, we have also studied the same system using cluster dynamical mean field theory (CDMFT)~\cite{li.ka.00,ko.sa.01,li.is.08,sene.15} with an exact diagonalization solver at zero temperature (or ED-CDMFT).
Here the Weiss fields of VCA are replaced by a bath of uncorrelated orbitals whose parameters are determined self-consistently.
Because the presence of this bath increases the size of the problem, the cluster cannot be as large as in VCA and typically contains no more than 4 sites. 

Each cluster, together with the associated bath, defines an Anderson impurity model (AIM):
\begin{equation}\label{eq:Himp}
H_{\rm imp} = H_c + \sum_{\mu,\a} \theta_{\mu,\a} \left(c_\mu^\dg a_\a^\fdg + \mbox{H.c.} \right)
+ \sum_{\a\b} \epsilon_{\a\b} a_\a^\dg a_\b^\fdg~~,
\end{equation}
where $a_\a$ annihilates an electron in the bath orbital labeled $\a$.
The Nambu formalism must be used to incorporate pairing between bath sites, within the matrix $\epsilon_{\a\b}$, or within the hybridization 
$\theta_{\mu,\a}$, depending on the impurity model.
The index $\mu$ then labels different sites of the cluster, together with the Nambu index, and takes $2L$ values in a cluster with $L$ sites.

\begin{figure*}[thp]
\includegraphics[width=0.9\hsize]{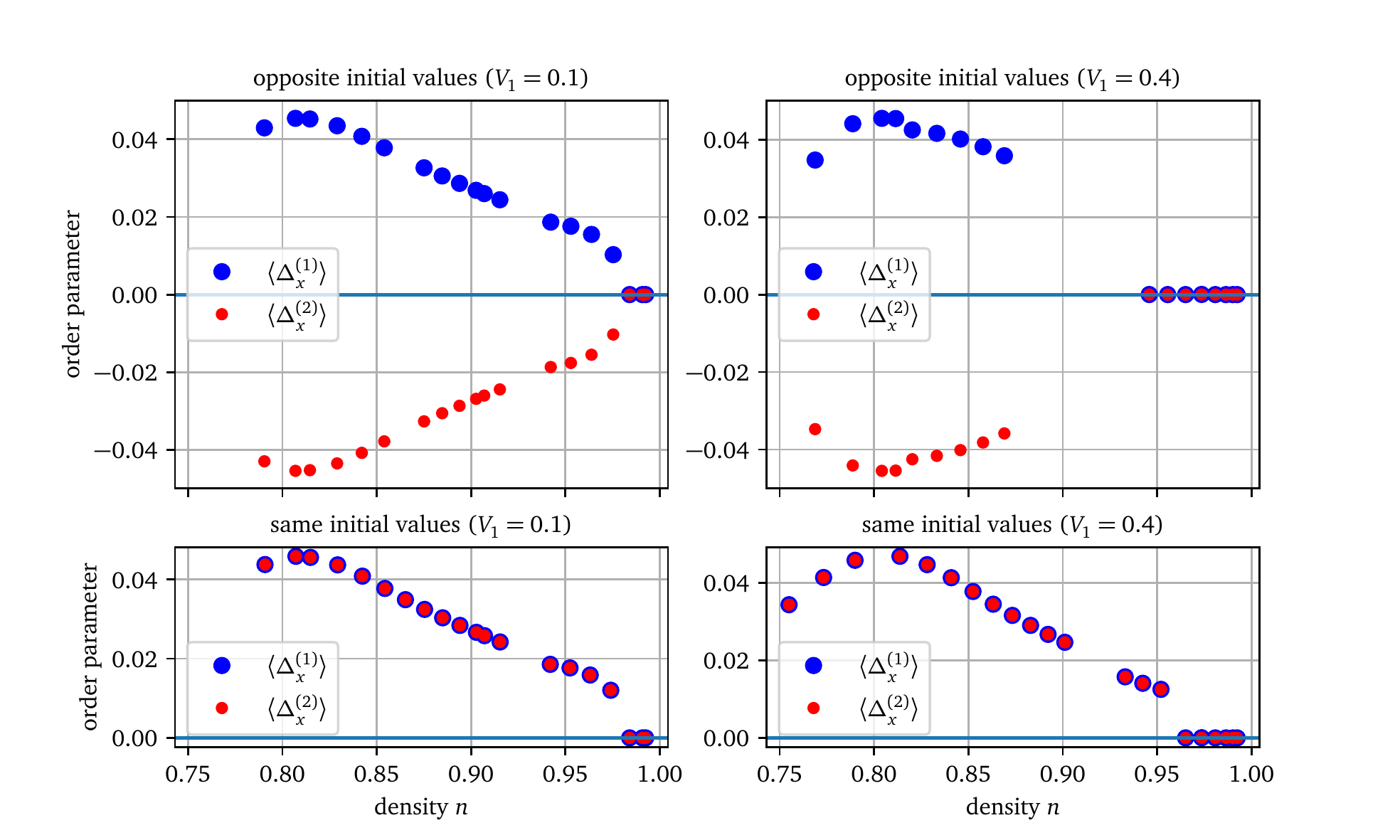}
\caption{(Color online) Order parameter as a function of electron density $n$, as found in CDMFT, for the two sets of inter-layer hopping. The anomalous bath parameters were initially set to have opposite (top) or identical (bottom) signs.}
\label{fig:cdmft}
\end{figure*}

The bath parameters $\theta_{\mu,\a}$ and $\epsilon_{\a\b}$ are determined by an approximate self-consistent procedure, as proposed initially in~\cite{ca.kr.94}, that goes as follows: (i) initial values of these parameters are chosen on the first iteration.
(ii) For each iteration, the cluster Hamiltonian \eqref{eq:Himp} is solved, i.e., the cluster Green function $\Gv_c(\omega)$ is computed.
The latter can be expressed as
\begin{equation}
\Gv_c(\omega)^{-1} = \omega - \tv_c - \Gammav(\omega) - \Sigmav_c(\omega)
\end{equation}
where $\Gammav(\omega)$ is the bath hybridization matrix:
\begin{equation}
\Gamma_{ij}(\omega) = \sum_{\a,\a'}\theta_{i\a}\left(\frac{1}{\omega - \epsilonv}\right)_{\a\a'}\theta_{j\a'}^*
\end{equation}
(iii) The bath parameters are updated, by minimizing the distance function:
\begin{equation}
d(\epsilonv, \thetav) = \sum_{i\omega_n} W(i\omega_n) \left[ \Gv_c(i\omega_n)^{-1} - \bar\Gv(i\omega_n)^{-1} \right]
\end{equation}
where $\bar\Gv(\omega)$, the projected Green function, is defined as
\begin{equation}\label{eq:GF}
\bar\Gv(\omega) = \frac{1}{N}\sum_\kv \Gv(\kv,\omega) \quad,\quad
\Gv(\kv,\omega) = \frac1{\omega - \tv_\kv - \Sigmav_c(\omega)}~~.
\end{equation}
Ideally, $\bar\Gv(\omega)$ should coincide with the impurity Green function $\Gv_c(\omega)$, but the finite number of bath parameters does not allow for this correspondence at all frequencies, and so a distance function $d(\epsilon_r, \theta_{ir})$ is defined, with emphasis on low frequencies along the imaginary axis.
The weight function $W(i\omega_n)$ is where the method has some arbitrariness; in this work $W(i\omega_n)$ is taken to be a constant for all Matsubara frequencies lower than a cutoff $\omega_c=2t$, with a fictitious temperature $\beta^{-1} = t/50$.
(iv) We go back to step (ii) and iterate until the bath parameters or the bath hybridization function $\Gammav(\omega)$ stop varying within some preset tolerance. 

In the current problem, the 10-site unit cell was separated in three impurity problems: a four-site cluster on each layer (which together are equivalent to the 8-site cluster used in VCA in the last section), made respectively of the orbitals (2,3,4,5) and (7,8,9,10) as labeled on Fig.~\ref{fig:cluster}, and a two-site cluster made of orbitals (1,6).
These clusters are illustrated on Fig.~\ref{fig:bath_simple_SC}.
The 4-site cluster is connected to 8 uncorrelated bath orbitals, and contains 6 independent parameters: Two bath energies $\epsilon_{1,2}$, two hybridization $\theta_{1,2}$ and two pairing amplitudes $\Delta_{1,2}$ between bath orbitals, with signs appropriate for describing $d$-wave superconductivity.
This way of parametrizing the bath is not the most general possible, but has been successfully used in the past~\cite{kancharla_anomalous_2008, kyung_pairing_2009,foley_coexistence_2019}.
The two-site cluster connects the two layers and also contains 6 bath parameters, except that the anomalous part is contained in the hybridization, i.e., it connects the bath sites to the cluster sites, not the bath sites themselves.
In order to allow for a relative phase between the pairing on the two layers, the pairing bath parameters $\Delta_{1,2}$ on the square cluster of the second layer are allowed to take complex values, whereas those on the first layer are assumed to be real.
Once a converged CDMFT solution is found, the same order parameters $\Delta^{(\ell)}$ as in the previous section are computed.

Fig.~\ref{fig:cdmft} shows the results of CDMFT applied to this system, for both intermediate (left) and strong (right) interlayer tunneling. 
The results depend on the initial set of bath parameters. 
On the top panels, the bath pairing parameters were initialized with opposite values on the two layers, whereas on the bottom panels, they were initialized with the same values.
At intermediate interlayer tunneling ($V_1=0.1$), the order parameters stay opposite throughout the doping range if the bath pairings are initialized this way; in other words, if the system is primed in the $B_2$ representation, it will stay in that representation.
At strong interlayer tunneling ($V_1=0.4$), this only occurs if doping is large enough. In other words, for doping 12\% or less, the system primed in the $B_2$ representation will either not converge, or converge to a normal solution, indicating its incompatibility with the $B_2$ initial conditions.
On the other hand, if the system is primed in the $B_1$ representation, then it stays in the $B_1$ representation, except that, at strong interlayer tunneling, it converges for larger values of doping, and converges to a normal solution at very small doping.

It is thus difficult to discriminate between the $B_1$ and $B_2$ representations within CDMFT, which does not have the fine energy resolution that VCA has. 
Nevertheless, we sense from the above results that the $B_1$ representation is preferred at low doping and the $B_2$ representation at higher doping, but a strong interlayer tunneling is needed for that.
Also, despite allowing in principle for an arbitrary complex phase between the anomalous bath parameters of the two layers, only the phases 0 and $\pi$ are found: no state with spontaneous breaking of time reversal is found in CDMFT.
This may be related to the fact that the main 4-site impurity model in CDMFT is confined to each layer, i.e., the complex, twisted inter-layer structure has an impact only through the self-consistency relation. In studying such systems, it seems that the VCA is a better choice.

\section{Conclusion}

In a one-band Hubbard model for a cuprate bilayer twisted by an angle of 53.13$^\circ$, the relative phase of the superconducting order parameter in the two layers depends on hole doping away from half-filling. In the underdoped regime, the relative phase vanishes, whereas it is $\pi$ in the overdoped regime. If the interlayer tunneling is strong, then there is an intermediate phase between those two in which this phase varies continuously from 0 to $\pi$. Time reversal symmetry is broken in that intermediate phase, but the topology is trivial, at least as computed from the electron Green function. At intermediate interlayer tunneling, this TRS breaking phase does not exist.

It is possible that this TRS breaking phase survives at weaker interlayer tunneling if the twist angle is closer to 45$^\circ$. A twist angle of $43.60^\circ$ corresponds to a unit cell of 58 copper sites
\cite{ca.tu.21} and might be amenable to a similar VCA study, albeit markedly more complex numerically. Work in this direction will be necessary in order to assess whether this putative phase is realistic in strong-coupling superconductivity. 

\begin{acknowledgments}
Computing resources were provided by Compute Canada and Calcul Qu\'ebec.
X.L. is supported by the National Natural Science Foundation of China (Grant No. 11974293) and the Fundamental Research Funds for Central Universities (Grant No. 20720180015).
D.S. acknowledges NSERC (Canada) under grant RGPIN-2020-05060.
\end{acknowledgments}


%

\end{document}